\documentclass[preprint,showpacs,preprintnumbers,amsmath,amssymb]{revtex4}

\usepackage{graphicx}
\usepackage{dcolumn}
\usepackage{bm}

\begin{document}

\preprint{APS/123-QED}

\title{Intra-granular $Ti$ precipitates in $MgB_2$ }

\author{M. J. Kramer}
\author{S. L. Bud'ko}%
\author{P. C. Canfield}%
\author{R. D. Wilke}%
\author{D. K. Finnemore}%
 \email{finnemor@ameslab.gov}
\affiliation{%
Ames Laboratory and Department of Physics and Astronomy\\
Iowa State University, Ames, IA 50011
}%
\author{Raymond J. Suplinskas}
\affiliation{%
Specialty Materials, Inc.\\
1449 Middlesex Street, Lowell MA 01851
}%

\date{\today}

\begin{abstract}
Superconducting magnesium diboride has been prepared with
intra-granular $TiB$ precipitates that have appropriate size and
spacing for the pinning of superconducting vortices. The $TiB$
particles have a size ranging from $1~nm$ to $20~nm$ and a spacing
about $5$ times as large.  By co-depositing $Ti$ and $B$ on a
commercial carbon coated silicon carbide substrate with chemical
vapor deposition methods, it was possible to obtain a starting
fiber with $Ti$ well dispersed in the $B$.  Reacting these fibers
in a $Mg$ vapor then transformed the fiber into a superconductor
capable of carrying $10,000~A/cm^2$ at $25~K$ and $1~T$.  The
$Ti$-$B$ bonds are strong enough that there is relatively little
migration of the $Ti$ as the $MgB_2$ grains form around the $TiB$
precipitates. $TiB_2$ precipitates, if they are present, can not
be seen in the selected area diffraction pictures that show the
$TiB$ precipitates.
\end{abstract}



\maketitle


In an important series of papers, Zhao and co-workers\cite
{1}showed that the critical current density, $J_c$, of $MgB_2$
could be substantially enhanced by $Ti$ doping.  Using a solid
state reaction of ground powders of $Mg$, $Ti$, and $B$, they
produced sintered pellets with $J_c~\sim 5 \times 10^4~A/cm^2$ at
$5~K$ and in $5~T$.  These were extremely small grained materials
with $MgB_2$ grains approximately $10~nm$ in size having $TiB_2$
decorating the grain boundaries.  High resolution transmission
electron microscopy seemed to indicate that the $TiB_2$ was only
one unit cell thick on the surface of the $MgB_2$ grains.  No $Ti$
was found within the $MgB_2$ grains.  It was suggested that the
presence of the $TiB_2$ limits the grain size of the $MgB_2$.  For
optimum $J_c$, approximately $10\% $ $Ti$ replacing $Mg$ is
needed.

Samples of  $Ti$-doped $MgB_2$ also can be prepared by a rather
different chemical vapor deposition (CVD) approach.\cite {2} In
this CVD deposition, $TiCl_4$ is added to the $BCl_3$ and $H_2$
gas stream and the $Ti$-doped $B$ is deposited on the conducting
substrate at about $1100^{\circ }C$.\cite {3} These $Ti$-doped
fibers are subsequently placed in a $Ta$ tube along with excess
$Mg$ and the fibers are transformed to $MgB_2$ in the $Mg$
vapor.\cite {4}  These conductors typically show $MgB_2$ grains in
range of $0.5~\mu m$ to $5~\mu m$ and critical current densities
of $10,000~A/cm^2$ at $25~K$ and $1.4~T$.\cite {5}.  These
enhancements are similar to those observed for $SiC$
nano-particles in $MgB_2$.\cite {6}

The purpose of this work is to report a transmission electron
microscope ($TEM$) study of the titanium boride precipitates in
the $MgB_2$ matrix. Scanning electron microscope ($SEM$)
pictures\cite {3} show that the precipitates are small compared to
the $1~\mu m$ $SEM$ beam size, and they show no evidence of
$TiB_2$ precipitates on grain boundaries as found for samples made
by solid state diffusion from ground powders of the elements.\cite
{1}  The goal here is to determine the size and distribution of
the titanium boride precipitates and to look for intra-granular
precipitates if they occur.

Titanium doped boron fibers are prepared in a reel-to-reel
continuous flow $CVD$ apparatus in which both the $B$ and $Ti$ are
deposited onto a $75~\mu m$ diameter $SiC$ fiber that has been
coated with a few micrometers of glassy carbon (commercially
available as $SCS-9A$).  The reactor is similar to those used to
make commercial boron fiber.  The reactant gas was predominantly
$BCl_3$ and $H_2$ gas at atmospheric pressure.  The $H_2$ gas is
bubbled through liquid $TiCl_4$ at zero degrees centigrade to give
a $TiC_4$ partial pressure of about $2\times 10^{-3}~bar$.  The
substrate is drawn through the reactor at about $100~mm/s$.  The
filament in the reactor was resistively heated to a peak
temperature of about $1100^{\circ }C$.  The resulting $Ti$-doped
$B$ layer was about $10~\mu m$ thick. To form $MgB_2$, short
lengths of fiber are places in a $Ta$ tube with about three times
excess $Mg$ and sealed under $Ar$. This tube in turn was sealed in
quartz and placed in a furnace at $950^{\circ }C$ for $2~h$.\cite
{4} Scanning electron microscope chemical analysis with energy
dispersive spectroscopy indicated variations in the $Ti/Mg$ ratio
of about three to five percent when a beam size of  a $1~\mu m$
diameter scans the sample.

To form a transmission electron microscope ($TEM$) sample, pieces
of the $Ti$-doped $MgB_2$ layer were flaked off the $SiC$
substrate and crushed. The resulting powder was floated onto a
holey carbon grid and grains having a thickness in the range of
$100$ to $200~nm$ were selected for study using a Philips $CM30$
$TEM$ operating at $300~keV$.

A typical grain of $Ti$-doped $MgB_2$ is shown in Fig. 1. From the
conditions of diffraction contrast in the $TEM$, the thickness of
the $MgB_2$ grain is estimated to be about $100$ to $200~nm$.
Typical precipitate size is $1$ to $10~nm$.  Because the
micrograph shows all the precipitates through the entire
thickness, the spacing of the overlapping precipitates is
substantially larger than the apparent separation of this image
due to the through thickness projection and is estimated to be in
the $20$ to $50~nm$ range. If an energy dispersive spectrum (EDS)
is taken over a large volume relative to the precipitate size, the
$Ti/Mg$ ratio is about $0.05$ in accord with the values found with
the $SEM/EDS$.

Tilting the grain so that a selected area diffraction pattern is
taken with the beam covering an approximately $500~nm$ portion of
the grain, a single hexagonal pattern is seen that indexes to be
$MgB_2$ as shown by Fig. 2. The ring patterns are presumably
caused by the precipitates and would indicate that the
precipitates are randomly oriented.  The crystal structure of
$TiB_2$ is very close to $MgB_2$ with nearly the same lattice
constant.  Hence the smaller rings probably do not arise from
$TiB_2$.

If the beam is tilted off the c-axis a few degrees, the ring
structure is more easily seen as shown by Fig. 3. The three most
prominent rings index well for the (111), the (200), and the (220)
lines of $TiB$.  If there is $TiB_2$ present, it is too small in
volume to be observed here. This is to be contrasted with the work
of Zhao and coworkers.\cite {1} They found that samples prepared
by solid state reaction of powders of $Ti$, $Mg$, and $B$ gave
$TiB_2$ precipitates on the grain boundaries of the $MgB_2$ grains
and no intra-granular precipitates.  It was a surprise to find the
dominant intra-granular precipitate to be $TiB$.  These results
indicate that the introduction of $Ti$ impurities into the $B$
before the $Mg$ is introduced, leads to a final morphology and
type of the $Ti$ precipitates different from the precipitates
found from solid state reaction of ground powders.\cite {1}

Critical current performance of these materials produced by $CVD$
are illustrated in Fig. 4.  At $5~K$, $J_c$ derived from
magnetization measurements\cite {4} drop from about $4\times
10^6~A/cm^2$ in self field to about $50,000~A/cm^2$ at $5~T$.  At
$25~K$, $J_c$ drops from about $900,000~A/cm^2$ in self field to
about $1,000~A/cm^2$ at $1.5~T$. For this material, the Meissner
screening in $50~Oe$ drops from zero at $36~K$ to $70~G$ at
$32~K$, so the effective $T_c$ has been suppressed by about $5~K$.

In summary, $Ti$-doped $MgB_2$ can prepared with a fairly uniform
array of intra-granular precipitates that is well suited for flux
pinning and high critical current densities.  The $Ti$ and $B$ can
be co-deposited by $CVD$ methods that are similar to those used to
make commercial boron fiber.  Then, the material can be converted
to $MgB_2$ in $Mg$ vapor.  The resulting $Ti$-doped $MgB_2$ is
found to have a dense array of randomly oriented $TiB$
precipitates have dimensions of about $1$ to $10~nm$ and a
separation $10$ to $50~nm$.  Critical current densities are
substantially enhanced over the pure $MgB_2$ material.

\begin{acknowledgments}
Ames Laboratory is operated for the US Department of Energy by
Iowa State University under Contract No. W-7405-Eng-82. This work
was supported by the Director for Energy Research, Office of Basic
Energy Sciences.
\end{acknowledgments}
\eject

\eject

\begin {figure}

\caption { $TEM$ micrograph of a single grain of $MgB2$ that is
about $100$ to $200~nm$ thick.}

\end {figure}

\begin {figure}

\caption { $TEM$ diffraction pattern for a large area of the grain
indicating a hexagonal pattern that indexes to $MgB_2$. The ring
patterns presumably arise from the precipitates. }

\end {figure}

\begin {figure}

\caption { Off axis $TEM$ diffraction pattern makes the ring
pattern easier to see.  The three most prominent rings index for
$TiB$.  }

\end {figure}

\begin {figure}

\caption { Critical current density for $Ti$-doped
$MgB_2$.
}

\end {figure}

\eject


\begin {references}

\bibitem{1} Y. Zhao, Y. Feng, C. H. Cheng, L. Zhou, Y. Wu, T. Machi, Y Fudamoto,
N. Koshizuka, and M. Murakami, Appl. Phys. Lett. {\bf 79}, 1154
(2001); Y. Zhao, D. X. Huang, Y. Feng, C. H. Cheng, T. Machi, N.
Koshizuka, and M. Murakami, Appl. Phys. Lett. {\bf 80} (2002)
1640.
\bibitem {2} R. J. Suplinskas and J. V. Marzik, Boron and silicon carbide filaments,
in: J. V. Milewski and H. S. Katz (Eds.)  \underline {Handbook of
Reinforcements for Plastics}, Van Nostrand Reinhold, New York,
1987.
\bibitem {3} D. K. Finnemore, W. E. Straszheim, S. L. Bud'ko, P. C. Canfield, N. E. Anderson,
Jr., and R. J. Suplinskas,  Physica C {\bf 385}, 278 (2003).
\bibitem {4}  P. C. Canfield, D. K. Finnemore, S. L. Bud'ko, J. E. Ostenson, G. Lapertot,
C. E. Cunningham, and C. Petrovic, Phys. Rev. Lett. {\bf 86}
(2001) 2423.
\bibitem {5} N. E. Anderson, W. E. Straszheim, S. L. Bud'ko, P. C. Canfield, D. K. Finnemore,
and R. J. Suplinskas, Physica C (submitted)
\bibitem {6}  S. X. Dou, S. Soltanian, J. Horvat, X. L. Wang, Z.
H. Zhou, M. Ionescu, H. K. Liu, P. Munroe, and M. Tomsic, Appl.
Phys. Lett. {\bf 81}, 3419 (2002).
\end {references}
\end{document}